\documentclass[reprint,twocolumn,balancelastpage]{revtex4} 
\usepackage{epsfig,subfigure,color,amsmath}

\begin{document} \title{A stochastic model of cascades in 2D turbulence}

\author{Peter D. Ditlevsen and Jes Ravnb{\o}l}
\affiliation{Centre for Ice and Climate, The Niels
Bohr Institute, University of Copenhagen
}

\date{\today} \maketitle

{\bf The dual cascade of energy and enstrophy in 2D turbulence cannot easily be understood
in terms of an analog to the Richardson-Kolmogorov scenario describing the energy cascade in 3D turbulence. The
coherent up- and downscale fluxes points to non-locality of interactions in spectral space, and thus the specific spatial structure
of the flow could be important. Shell models, which lack spacial structure and have only local interactions in spectral space, 
indeed fail in reproducing the correct scaling for the inverse cascade of energy. In order to exclude the possibility that
non-locality of interactions in spectral space is crucial for the dual cascade, we introduce a stochastic spectral model of the cascades
which is local in spectral space and which shows the correct scaling for both the direct enstrophy - and the inverse energy
cascade.}

The scaling relations for the energy spectrum in turbulence are consequences of the energy cascade in 3D turbulence  \cite{kolmogorov:1941a} and the dual cascades of energy and enstrophy in 2D turbulence  \cite{kraichnan:1967}. The cascades follow from inviscid conservation of energy and enstrophy (in 2D), separation between forcing and dissipation scales and the sweeping of smaller eddies by larger scale flow, showing that interactions are local in spectral space. The last assumption is based on the Richardson picture of energy flowing in a self-similar manner from large to small scales \cite{frisch:1995}, and thus depends somewhat on the actual physical structure of the flow. In fact, experiments  \cite{douady:1991} and simulations \cite{jimenez:1993} show that fully developed 3D turbulence has a filamented structure rather than a structure with eddies, or vortices, with smaller eddies inside as
envisaged by Richardson. 

In 2D turbulence the situation is quite different: Since enstropy is also an inviscid invariant, the energy cascade is from smaller to larger scales.  This inverse energy cascade is seen in both experiments \cite{paret:1997} and simulations \cite{boffetta:2010,vallgren:2011}. The scenario fits poorly into the Richardson picture. The 2D turbulent flow is, besides the lower dimensionality also different in its spatial structures from 3D turbulent flow. A 2D turbulence analog for the Richardson picture could be as follows: The flow is characterized by well localized energy-containing vortices where the flow in between the vortices is characterized by a strong shear accounting for the enstrophy dissipation. Vortices can merge, leading to even larger scale structures. For this picture the dual cascade can be explained in a simple heuristic scenario:
Consider a vortex of linear scale $R$ rotating as a rigid body  with 
rotational speed $\Omega$. Thus the velocity is $u_i({\bf r})=\epsilon_{ijl}\Omega_jr_l$ for $r<R$, falling off
rapidly for $r>R$. $\Omega_i$ is perpendicular to the plane of the flow. The energy of the vortex
is $E=(\pi/4)\Omega^2R^4$ and the entrophy is $Z=4\pi \Omega^2R^2$.

\begin{figure}[h] 
\epsfxsize=9cm 
\epsffile{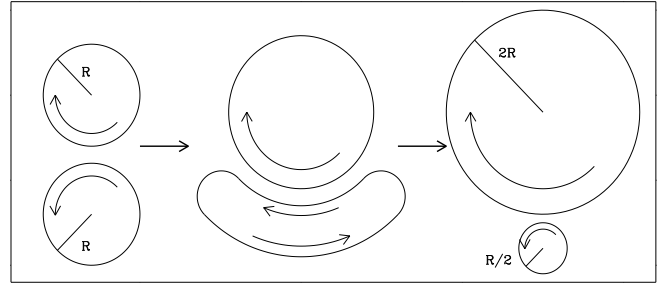}\caption[]{Schematic cartoon of the scattering of two vortices of radii $R$ in a 2D flow. 
The lower vortex is stretched in the flow of the upper vortex. This results in growth to size 
$2R$
of the upper vortex while a smaller vortex of size $R/2$ is scattered of. If the vortices
are considered to perform rigid body rotations, the big upper vortex contains most of 
the energy while the
small lower vortex contains most of the enstrophy. In this way the energy is cascaded to large
scales while enstrophy is cascaded to small scales.
}
\label{fig:vort}
\end{figure}

Consider now a flow of two such vortices of linear size $R$. Assume that they scatter in process after which
two vortices of linear sizes, say,
$R/2$ and $2R$ emerge. This is schematically shown in figure \ref{fig:vort}. 
These new vortices have rotational speeds $\Omega_1$ and
$\Omega_2$ respectively. From energy and enstrophy conservation $\Omega_1$ and
$\Omega_2$ are determined,
\begin{align}
E/(\pi/4)=2\Omega^2R^4&=\Omega_1^2(R/2)^4+\Omega_2^2(2R)^4 \nonumber \\
Z/(4\pi)=2 \Omega^2 R^2&= \Omega_1^2 (R/2)^2 +  \Omega_2^2 (2R)^2
\end{align}
from which we get $\Omega_1^2=32\Omega^2/5$ and $\Omega_2^2=\Omega^2/10$.
The energy is then redistributed such that $E_1=E/5$ and $E_2=4E/5$,
while the enstrophy is distributed such that
$Z_1=4Z/5$ and $Z_2=Z/5$.
Thus the energy has moved to larger scales while the enstrophy has moved to smaller scales. 
The role of coherent structures and scattering of vortices in the cascade process is not at present clear and this is just one
of several conceptual pictures proposed \cite{tabeling:2002,chen:2006}.

A complementary approach to modeling the cascade process  is taken in the one dimensional shell models introduced by Obukhov \cite{obukhov:1971} and Gledzer \cite{gledzer:1973}.
In these models there are no meaningful representation of the spatial structure of the flow.
The flow is represented by a set of generalized spectral velocity components $u_n$,
associated with a wave number $k_n=\lambda^n$, where $\lambda$ is a spectral shell spacing (typically $\lambda=2$).
The velocity $u_n$ can be interpreted as some average representation of all spectral fluid velocity components $u(k)$ within
a shell $k_{n-1}<|k|<k_n$, thus the name 'shell model'. The dynamics of the shell models  \cite{lvov:1998,ditlevsen:2000a}  are, except for the tensorial structure, similar to the spectral Navier-Stokes equation: 

\begin{align}
\dot{u}_n&=i \,k_n \left(u_{n+1}^*u_{n+2}-\frac{\epsilon}{\lambda} u_{n-
1}^*u_{n+1} -\frac{\epsilon-1}{\lambda^2} u_{n-2}u_{n-1}\right)
\nonumber \\ 
&-(\nu k_n^2+\nu_1k_n^{-2}) u_n +f\delta_{n,n_0}
\label{sabrae}
\end{align}
The terms in the first parenthesis correspond to the non-linear advection and pressure gradient terms. The next term is the
viscous dissipation and in the 2D case the large scale drag. The last term is a forcing term localized at some wave number $k_{n_0}$.  
The shell models have two quadratic inviscid invariants, energy: 

\begin{equation}
E=\sum_n |u_n|^2,
\end{equation}
and a second invariant determined by the free parameter $\epsilon$:
\begin{equation}
E_2=\sum_n (\epsilon-1)^{-n} |u_n|^2=\sum_n k_n^\alpha |u_n|^2,
\end{equation}
where the last equality defines the exponent $\alpha=-\log(\epsilon-1)/\log\lambda$. 
For $0<\epsilon<1$ the factor $(\epsilon-1)^{-n}=(-1)^n|\epsilon-1|^{-n}$ has alternating signs for even and odd 
shell number $n$ corresponding to a generalized helicity  \cite{kadanoff:1997,ditlevsen:1997}. For these parameter values the shell models are denoted 3D-like. For $1<\epsilon<2$ the second invariant is always positive corresponding to a generalized enstrophy, and the models are denoted 2D-like.
The enstrophy has the same dimension as in the real 2D flow for $\alpha=2$ and thus $\epsilon=5/4$.
The velocity in the shell models have no meaningful spatial structure,
but the 3D-like shell models do exhibit a forward energy cascade, with a Kolmogorov scaling (K41) relation $\langle |u_n|\rangle\sim k_n^{-1/3}$. 
Recent interest in the 3D-like shell models has been on the numerical finding that not only do the models show K41 scaling relations, the models also show 
intermittency corrections to K41 leading to anomalous scaling relations similar to what is seen in high Reynolds number 3D turbulence \cite{crisanti:1994,anselmet:1984}. 
From inviscid energy conservation, there is an exact scaling
relation for the non-linear flux of energy $\langle{\Pi}_n\rangle=k_{n}\Delta_{n+1}-k_{n-1}(\epsilon-1)\Delta_n
=\overline{\varepsilon}$, where $\overline{\varepsilon}$ is the mean energy
dissipation, and $\Delta_n=\langle{\it Im}(u_{n-1}u_nu_{n+1})\rangle$ is a specific third order structure function. 
This corresponds to the 4/5th law of homogeneous and isotropic 3D turbulence.

For the rest of this paper we shall focus on the 2D-like models and denote $E_2=Z$.
This case is more tricky: 
From classical scaling arguments we get the (constant) mean non-linear flux of enstrophy through the inertial range
as  $\langle{\Pi}_n^Z\rangle \sim k_n^{\alpha+1}\langle |u_n|\rangle^3 \Rightarrow \langle |u_n|\rangle \sim k_n^{-(\alpha +1)/3}$. This is the corresponding Kolmogorov-Kraichnan scaling for the shell models. 
Obviously, in this case, as in 2D turbulence, the inviscid enstrophy conservation also lead to an exact scaling relation for a specific third order structure function \cite{tabeling:2002}. 
One heuristic argument for the transfer of enstrophy to smaller scales (larger wave numbers) is that the wave-wave interactions
will tend to distribute enstrophy evenly over the degrees of freedom of the system, which is the maximum entropy state. 
This state of equipartition of enstrophy defines a different scaling relation; $k_n^\alpha \langle |u_n|^2\rangle\sim \text{const.} \Rightarrow \langle |u_n|\rangle\sim k_n^{-\alpha/2}$.

Now, for the dimensionally correct enstrophy $(\alpha=2)$ the two scalings are the same, so a cascade and
a diffusive transport of enstrophy in quasi equilibrium cannot be distinguished \cite{aurell:1994}. 
This is an artifact of the shell models
not present in 2D turbulence, where the spectral slope for enstrophy cascade is $\langle |u(k)|\rangle\sim k^{-1}$, while for equipartition it is $\langle |u(k)|\rangle \sim k^{-1/2}$. If the exponent $\alpha$ is different from 2, the scalings corresponding to enstrophy cascade and equipartition
are different. 
For $0<\alpha \le 2$ the 2D-like shell models show a forward enstrophy cascade, while for $\alpha \ge 2$ they
show an equipartitioning of enstrophy. This numerical finding could be related to how the typical eddy turnover time $\tau_n$ depends on wave number. The typical eddy turnover time is simply defined by dimensional counting: $\tau_n=(k_n\langle |u_n|\rangle)^{-1}=(k_n\sqrt{\langle E_n\rangle})^{-1}$.  
Assuming the scaling $\overline{E}_n\sim k_n^\gamma$, the scaling for the eddy turnover time becomes $\tau_n \sim k_n^\kappa=k_n^{-(\gamma+2)/2}$.
The scaling exponents in the energy range and for three values of $\alpha$ in enstrophy range in the two cases of cascade or equipartition and are summarized in Table 1.

\begin{table}[ht]
\begin{tabular}{|c||c|c||c|c|}
\hline 
 & \multicolumn{2}{c||}{\hspace{3mm}Cascade\hspace{3mm}   } &  \multicolumn{2}{c|}{Equipartition}  \\
\hline
\hspace{2mm}\hspace{2mm} & \hspace{3mm}$\gamma$\hspace{3mm} &  $\kappa$ &  \hspace{2mm} $\gamma$ \hspace{2mm}  &  $\kappa$ \\
\hline
Energy          &    -2/3      &   -2/3    &  0     & -1  \\ 
\hline
$\alpha=1 $         &    -4/3      &    -1/3   &  -1    &  -1/2 \\  
\hspace{6mm}2          &   -2          &     0  &  -2    &   0 \\
\hspace{6mm}3          &   -8/3       &     1/3  & -3     &  1/2 \\
\hline
\end{tabular}
\caption{Scaling exponents $\gamma$ and $\kappa$ for the energy spectrum, $E_n\sim k_n^\gamma$, and the eddy turnover time, $\tau_n\sim k_n^\kappa$ in the cases
of cascade or equipartition of energy or enstrophy  respectively.}
\end{table}

\begin{figure}[h] \begin{center}\epsfxsize=9cm 
\epsffile{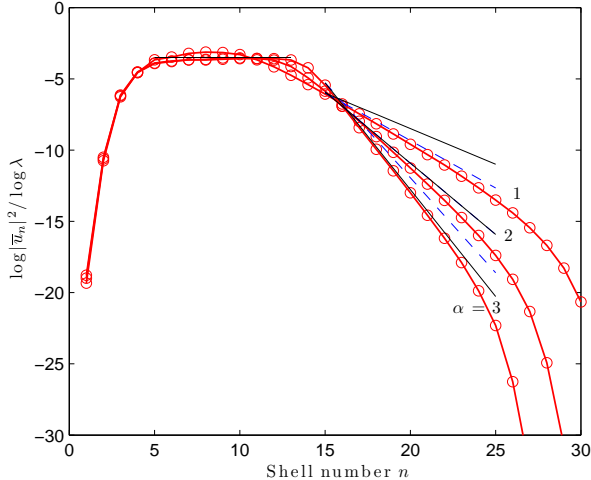}
\caption[]{Shell model energy spectra for $\alpha=1, 2, 3$, $(\epsilon=1+\lambda^{-\alpha})$. The pumping scale is at wave number $n_0=15$. Other parameters of the simulations are: $\nu=10^{-14}, \nu_1=100, f=1+i$. The dashed lines are the scaling relations corresponding to (forward) cascade of enstrophy , while the full lines corresponds to equipartition of enstrophy. When the slope corresponding to equipartition is steeper than the
slope corresponding to cascade ($\alpha=3$), the model shows equipartition, while in the opposite case ($\alpha=1$) it shows cascade. By the same token there is equipartition of energy in the inverse cascade range.  
}
\label{sabra_figure} \end{center}\end{figure}

In the case of a cascade ($0<\alpha \le 2$) the typical turnover time decrease with
increasing wave number, while in the case of equipartition ($\alpha\ge 2$) it increase with wave number, leaving 
time for upscale (from large to small wave numbers) transport of enstrophy to equilibrate \cite{ditlevsen:1996a,ditlevsen:2011}. For the same
reason the 2D-like shell models fail in simulating the inverse cascade of energy: For a spectrum corresponding to energy cascade the 
eddy turnover time decrease with increasing wave number (see Table 1), which makes the transport of energy diffusive, 
preventing the classical inverse energy cascade (independent of $\alpha$). 
The situation is summarized in figure \ref{sabra_figure}, where the energy spectra for the 
three cases, $\alpha= 1, 2, 3$ are shown: For $\alpha=1$ the model has a cascade spectrum (dashed line), for $\alpha=3$
it has an equilibrium spectrum (full line), while for $\alpha=2$ the two spectra coincide.

It thus seems that one dimensional models are unable to generate
the dual cascade phenomenon characteristic for 2D turbulence, which suggests that the specific spatial structure
of the flow is essential for the inverse energy cascade.
In order to investigate this or if the dual cascade is related
to specific scale dependence of turnover times we construct a stochastic Markov chain model of the cascade process.
As for the shell models we define a chain of exponentially growing scales in wave number space $k_n=\lambda^n$. The
dynamical variable associated with each scale is the energy $E_n$. The enstrophy $Z_n$ is related to the energy as
$Z_n=k_n^\alpha E_n$. The stochastic dynamical equation for $E_n$ is:
\begin{eqnarray}
dE_n=&\{&q_{n+1}E_{n+1}+(\epsilon-1)q_{n-1}E_{n-1}-\epsilon q_nE_n \nonumber \\
&+&\tilde{q}_{n+1}\tilde{E}_{n+1}+(\epsilon-1)\tilde{q}_{n-1}\tilde{E}_{n-1}-\epsilon \tilde{q}_n\tilde{E}_n \nonumber \\
&-&(\nu k_n^2+\nu_1 k_n^{-2}) E_n + f\delta_{n,n_0}\}\, dt
\end{eqnarray}
where $\tilde{E}_n=\sqrt{E_{n-1}E_{n+1}}$ and $\{ (q_n, \tilde{q}_n)$, $n=1, ... , N \}$ is a set of 2N stochastic variables:
\begin{align}
q_{n} \, (\tilde{q}_n)=& \left\{\begin{array}{rl} 
1/\delta\tau\,  (-1/\delta\tau)& \text{with probability } P_n\\
0 & \text{with probability }(1-P_n) \end{array}\right. \label{prjump}
\end{align}
where $P_n=\text{min}(1, \delta\tau/\tau_n)$, $\tau_n=1/(k_n\sqrt{E_n})$ is defined as a 
dynamical eddy turnover time and $\delta\tau$ is a time interval smaller than the smallest time scale in the system. It is straight forward to verify that energy and
enstrophy are conserved in the unforced and inviscid case. The case $q_n=1/\delta\tau$ corresponds to a triad interaction where energy is transferred 
from shell $n$ to shells $n-1$ and $n+1$. The case $\tilde{q}_n=-1/\delta\tau$ corresponds to a triade interaction 
where energy
is transferred from shells $n-1$ and $n+1$ to shell $n$. The choice of $\tilde{E}_n$ as the geometric mean of the energies of the neighboring shells ensures that energies remain positive. Furthermore, in the case that the energy follows
a perfect scaling relation, $E_n= E_0 (k_n/k_0)^\gamma$, we have $\tilde{E}_n=E_n$ and the model has detailed balance in the sense that a (positive) energy/enstrophy transfer from shell $n$ to the two neighboring shells has the same probability 
as a transfer in the opposite direction. 

\begin{figure}[h] \begin{center}\epsfxsize=9cm 
\epsffile{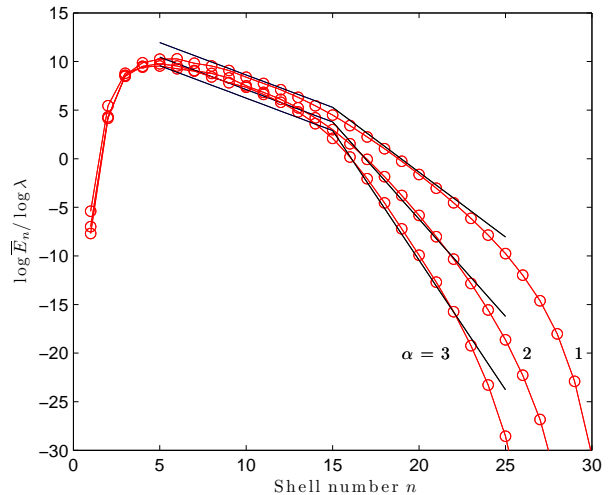}
\caption[]{Stochastic model energy spectra for $\alpha=1, 2, 3$, $(\epsilon=1+\lambda^{-\alpha})$. The pumping scale is at wave number $n_0=15$. Other parameters of the simulations are: $\nu=10^{-17}, \nu_1=0.5, f=0.1$.The lines for $n>n_0$ are the scaling relations corresponding to (forward) cascade of enstrophy, while the lines for $n<n_0$ correspond to the K41 scaling for the inverse cascade of energy.}
\label{spec_figure} \end{center}\end{figure}

The stochastic model energy spectra for the three cases $\alpha= 1, 2, 3$
are shown in figure \ref{spec_figure}. 
In all three cases the two scaling regimes of inverse cascade of energy and forward cascade of enstrophy are observed.
The stochastic model thus, in contrast to the shell models, shows the same behavior of dual cascade as in 2D 
turbulence, so even though the eddy turnover time in the spectral range of inverse cascade of energy decrease with
wave number, the system will not equilibrate. 
In the spectral range of inverse energy cascade where large scale energy dissipation and smaller scale energy pumping
is well separated, there is a statistical steady state, $\langle{\Pi}_n\rangle = \overline{\varepsilon}$, where again $\overline{\varepsilon}$ is the mean energy dissipation.
Correspondingly in the range of forward enstrophy cascade there is a statistical steady state, $\langle{\Pi}_n^Z\rangle = \overline{\eta}$, where $\overline{\eta}$ is the mean enstrophy dissipation.
Similar to the shell model, the mean non-linear transfer  of energy $\langle{\Pi}_n\rangle$ and enstrophy $\langle{\Pi}^Z_n\rangle$ from shells $m\le n$ to
shells $m>n$ are easily calculated: 
\begin{align}
\langle\Pi_n\rangle =\langle q_{n+1}E_{n+1}\rangle +(1-\epsilon)\langle q_{n}E_{n}\rangle &&\nonumber \\
+
\langle\tilde{q}_{n+1}\tilde{E}_{n+1}\rangle +(1-\epsilon)\langle\tilde{q}_{n}\tilde{E}_{n}\rangle&&
\label{piE}
\end{align}
and 
\begin{equation}
\langle\Pi^Z_n\rangle =
k_n^\alpha(\langle q_{n+1}E_{n+1}\rangle-\langle q_{n}E_{n}\rangle 
+
\langle\tilde{q}_{n+1}\tilde{E}_{n+1}\rangle -\langle\tilde{q}_{n}\tilde{E}_{n}\rangle ).
\label{piZ}
\end{equation} 
Each of the terms on the right hand sides has the form $\langle q_nE_n\rangle=\langle P_nE_n\rangle=k_n\langle E_n^{3/2}\rangle\equiv k_n\Delta_n$ or
$\langle \tilde{q}_n\tilde{E}_n\rangle=-\langle P_n\tilde{E}_n\rangle=-k_n\langle (E_{n-1}E_nE_{n+1})^{1/2}\rangle\equiv -k_n\tilde{\Delta}_n$ and equations
(\ref{piE}) and (\ref{piZ}) can be rewritten 
\begin{align}
\langle\Pi_n\rangle=k_{n}\{\lambda (\Delta_{n+1}-\tilde{\Delta}_{n+1})+(1-\epsilon)(\Delta_n-\tilde{\Delta}_n)\}
\label{piE1}
\end{align}
and  
\begin{align}
\langle\Pi_n^Z\rangle=k_{n}^{\alpha+1}\{\lambda(\Delta_{n+1}-\tilde{\Delta}_{n+1})-(\Delta_n-\tilde{\Delta}_n)\}.
\label{piZ1}
\end{align}
An exact scaling relation $\Delta_n=\tilde{\Delta}_n=c k_n^{3\gamma/2}$ would imply $\langle\Pi_n\rangle=\langle\Pi^Z_n\rangle=0$ violating the non-zero
inverse energy and forward enstrophy cascades. Numerical inspection shows that $(\Delta_n-\tilde{\Delta}_n)/\Delta_n \approx 0.02$ independent of $n$ (and
$\alpha$). Thus we may assume a K41 scaling relation $(\Delta_n-\tilde{\Delta}_n)=C k_n^{3\gamma/2}$. In the range of inverse energy cascade, $k_n<k_{n_f}$ 
(or forward enstrophy cascade, $k_n > k_{n_f}$),
$\langle\Pi_n\rangle=\overline{\varepsilon} \mbox { (or }   0)$ and $\langle\Pi_n^Z\rangle=0 \mbox{ (or } \overline{\eta})$, equations (\ref{piE1}) and (\ref{piZ1}) implies:
\begin{equation}
\langle\Pi_n\rangle=C k_n^{1+3\gamma/2}\{\lambda^{1+3\gamma/2}+(1-\epsilon)\}=\overline{\varepsilon}  \mbox{ (or } 0) 
\label{piE2}
\end{equation}
and
\begin{equation}
\langle\Pi_n^Z\rangle=C k_n^{\alpha+1+3\gamma/2}\{\lambda^{1+3\gamma/2}-1\}=0 \mbox{ (or } \overline{\eta}).
\label{piZ2}
\end{equation}

Both equations are fulfilled exactly when $\gamma=-2/3$ and $C=\overline{\varepsilon}/(2-\epsilon)$ (or in the case of enstrophy cascade, $\gamma=-2(\alpha+1)/3$ and $C=\overline{\eta}/(\epsilon-2)$), note that $\lambda^{-\alpha}=(\epsilon-1)$. The scaling solutions corresponding to the dual Kolmogorov-Kraichnan cascades are
obtained here from the exact cancellations of the two terms in the curly brackets.

In conclusion, the behavior of the stochastic model exhibiting dual cascade indicates that the scaling arguments leading to the prediction of dual cascade in 2D turbulence are indeed robust and that 
long range triade interactions in the spectral domain are not crucial for explaining the dual cascade. The model furthermore challenges
the suggestion that the reason for why shell models exhibit equilibrium spectra and fail in reproducing the Kolmogorov spectrum for the inverse energy cascade should be related to the typical eddy turnover time scales leaving time for the energy to equilibrate before
being cascaded up-scale. The difference in the behavior between the shell model and the stochastic model is rather connected to
the fact that the shell model, in contrast to the stochastic model, is quadratic in the invariants (energy and enstrophy) with respect to the dynamical variables (velocities). 
This implies that in the inviscid and unforced case, they obey the equipartition theorem, leading to quasi-equilibrium also in the 
case of forcing and dissipation.


\end{document}